\documentclass[preprint,review,sort&compress,12pt]{elsarticle}

\usepackage{tikz}
\usetikzlibrary{shapes}
\newcommand{\tikzcirclered}[2][black,fill=red]{\tikz[baseline=-0.5ex]\draw[#1,radius=#2] (0,0) circle ;}%
\newcommand{\tikzcirclegray}[2][gray!60,fill=gray!60]{\tikz[baseline=-0.5ex]\draw[#1,radius=#2] (0,0) circle ;}%

\usepackage{amssymb}
\usepackage{amsthm}
\usepackage{amsmath}
\usepackage{bm}
\usepackage{mathtools}
\usepackage{empheq}
\usepackage{graphicx}
\usepackage{caption}
\usepackage{subcaption}
\usepackage[font=normalsize,labelfont=bf]{caption}
\usepackage{algorithm}
\usepackage{algpseudocode}
\usepackage{array}
\usepackage{multirow}
\usepackage{enumerate}
\usepackage{diagbox} 
\usepackage{lineno}
\usepackage{fullpage}
\usepackage{xcolor}
\definecolor{ForestGreen}{RGB}{34,139,34}
\usepackage[colorinlistoftodos]{todonotes}
\usepackage{bbm}
\usepackage{listings}
\usepackage{url}
\usepackage{textcmds}

\graphicspath{ {./figs/} }
\biboptions{numbers,comma,round,square}

\graphicspath{ {./figs/} }

\journal{Journal of The Royal Society Interface}
\begin{document}
\begin{frontmatter}

 \title{Deep learning-based surrogate model for 3-D patient-specific computational fluid dynamics}




\author[ndAME]{Pan Du}
\author[ndACMS]{Xiaozhi Zhu}
\author[ndAME]{Jian-Xun Wang\corref{corxh}}

\address[ndAME]{Aerospace and Mechanical Engineering, University of Notre Dame, Notre Dame, IN}
\address[ndACMS]{Applied and Computational Mathematics and Statistics, University of Notre Dame, Notre Dame, IN}
\cortext[corxh]{Corresponding author. Tel: +1 574-631-5302}
\ead{jwang33@nd.edu}

\begin{abstract}
Optimization and uncertainty quantification have been playing an increasingly important role in computational hemodynamics. However, existing methods based on principled modeling and classic numerical techniques have faced significant challenges, particularly when it comes to complex 3D patient-specific shapes in the real world. First, it is notoriously challenging to parameterize the input space of arbitrarily complex 3-D geometries. Second, the process often involves massive forward simulations, which are extremely computationally demanding or even infeasible. We propose a novel deep learning surrogate modeling solution to address these challenges and enable rapid hemodynamic predictions. Specifically, a statistical generative model for 3-D patient-specific shapes is developed based on a small set of baseline patient-specific geometries. An unsupervised shape correspondence solution is used to enable geometric morphing and scalable shape synthesis statistically. Moreover, a simulation routine is developed for automatic data generation by automatic meshing, boundary setting, simulation, and post-processing. An efficient supervised learning solution is proposed to map the geometric inputs to the hemodynamics predictions in latent spaces. Numerical studies on aortic flows are conducted to demonstrate the effectiveness and merit of the proposed techniques.

\end{abstract}

\end{frontmatter}

\nolinenumbers

\section{Introduction}
\label{sec:intro}

Cardiovascular disease (CVD) remains the first leading cause of death and morbidity in the U.S., posing a major healthcare concern~\cite{mozaffarian2015executive}. Although medical imaging techniques, e.g., computed tomography (CT) and magnetic resonance imaging (MRI), have enabled the acquisition of exquisite anatomical information and revolutionized cardiovascular medicine~\cite{nayak2015cardiovascular}, they are usually not able to provide hemodynamic information (e.g., flow field, pressure losses) on their own, which is believed to be more important to CVD diagnosis and prognosis~\cite{taylor2009patient}. 
Computational models based on physical principles 
of cardiovascular systems, combined with medical imaging, enable the derivation of functional information inaccessible by medical images alone. In the past decade, image-based computational modeling has become a paradigm in cardiovascular research~\cite{steinman2005flow,xiang2016aview,quarteroni2017cardiovascular} and is pioneering new clinical applications~\cite{taylor2013computational,chinnaiyan2017rationale,hokken2020precision}.  
However, the first-principle physics-based models are often computationally expensive since they involve solving a large-scale discrete partial differential equation (PDE) system using numerical techniques, e.g., finite volume or finite element methods (FVM/FEM). Particularly, when studying dynamics of blood flows (i.e., hemodynamics), computational fluid dynamics (CFD) or/and fluid-structure interaction (FSI) models are needed, which are enormously time-consuming (requiring supercomputing clusters) and prone to numerical challenges (requiring significant domain expertise). These roadblocks have primarily limited image-based CFD modeling in clinical applications that require timely feedback for further therapeutic assessment and treatment planning. For example, in the case of Heartflow (a medical company for precision heart care), image data must be sent outside the hospital to be processed by in-house supercomputers and results sent back to the hospital. Likewise, the high computational cost prohibits many-query simulations for, e.g., uncertainty quantification, parameter estimation, design optimization, etc., which are becoming increasingly vital in advancing the utility of cardiovascular modeling to practical applications~\cite{sankaran2011stochastic,sankaran2010impact,tran2017automated,schiavazzi2017generalized,gao2021bi,gao2020bi1}.


As a more efficient alternative to the principled CFD/FSI models, reduced-order modeling (ROM) has been an area of intense investigation for years. One class of ROM is constructed by simplifying the physics of cardiovascular systems, such as lumped parameter (LP) models, one-dimensional (1-D) models, or hybrid 3-D/1-D/LP models~\cite{mirramezani2019reduced,mirramezani2020distributed,pfaller2021automated}, where the physical dimensions are reduced to different levels. For example, LP models typically treat the hemodynamic system as an electric circuit, where the relations among pressure drop, viscous effects, and flow velocity are modeled as simplified ordinary differential equations (ODEs). Although LP/1D models are very cheap, they only can predict global information (integral quantities) instead of local ones (e.g., spatiotemporal fields of velocity or wall shear stresses), while the latter is more useful to advance cardiovascular research/healthcare. To preserve detailed hemodynamics information, the other class of ROM is built by projecting the full-order governing PDEs (e.g., Navier-Stokes equations) onto a reduced subspace spanned by a group of basis functions, such as proper orthogonal decomposition (POD) modes, known as POD-Galerkin projection, which has been developed for facilitating hemodynamics analysis in many 3-D patient-specific configurations~\cite{manzoni2012model,lassila2013reduced,ballarin2016pod,ballarin2016fast}. However, the projection-based ROM is often less stable in parametric settings and highly code-intrusive, posing great challenges to leveraging legacy CFD/FSI solvers~\cite{lassila2014model,chaturantabut2010nonlinear,gao2020non}.    
With the increasing data availability and recent advances in machine learning (ML), there has been growing interest in developing non-intrusive data-driven approaches for surrogate modeling of cardiovascular systems. In general, data-driven surrogate models aim to learn the mapping between modeling inputs (e.g., geometry, inlet/outlet, material properties, etc.) and computed outputs (e.g., resolved flow field, pressure contour, wall shear stress distribution, etc.) based on full-fidelity CFD/FSI simulation data. A well-trained surrogate model can predict detailed hemodynamics information rapidly, supporting real-time or many-query applications. For example, collocation-based polynomial chaos expansion (PCE) has been used to build non-intrusive surrogates of 3-D blood flow simulations for UQ tasks~\cite{sankaran2010impact,sankaran2011stochastic}. Gao et al.~\cite{gao2021bi,gao2020bi1} developed a bi-fidelity surrogate model that leverages the efficiency and accuracy of the low- and high-fidelity CFD simulations, respectively. Deep neural network (DNN) has become a popular surrogate modeling approach, renowned for their universal functional approximation capability for high-dimensional system~\cite{scarselli1998universal,hutzenthaler2020overcoming}. Most recently, DNN-based surrogate models have been developed for cardiovascular applications and shown great potential~\cite{liang2017machine,liang2018deep,madani2019bridging,balu2019deep,sahli2020physics,kissas2020machine,liang2020feasibility,li2021prediction}.

However, ML-based surrogate modeling for complex patient-specific hemodynamics still faces significant challenges that existing techniques struggle to address simultaneously. First, the parameterization of the geometric space spanned by irregular 3-D patient-specific shapes is very challenging.  As a result, most existing ML-based surrogate hemodynamic models focus on inlet/outlet parameterization or studying idealized geometries, which can be easily parameterized with a few descriptors (e.g., radius), and only a few studies dealt with 3-D patient-specific cases~\cite{liang2020feasibility,li2021prediction}. 
Second, a well-performed ML surrogate model often requires a large number of training samples. However, patient-specific geometries are commonly obtained by segmentation from medical images (e.g., CT and MRA), which are often cumbersome and time-consuming, leading to a prohibitive cost for data generation. To address the issue, it is necessary to effectively synthesize numerous new samples from a small set of patient-specific geometries obtained by image segmentation.   

In this paper, we try to fill the gaps and develop an ML-based surrogate modeling framework for rapid predictions of comprehensive hemodynamics information in 3-D patient-specific aortic geometries.  
In contrast to many ML-based surrogates that predict global hemodynamics information such as pressure loss, FFR, and mean wall shear stress (WSS), this work focuses on predicting local blood flow information (e.g., velocity, pressure, and WSS fields), given irregular 3-D patient-specific shapes. The contributions of the current paper are summarized as follows. First, we developed a method to parameterize 3-D patient-specific aorta geometries based on stochastic shape modeling (SSM). Second, a generative model is proposed to synthesize a massive amount of 3-D geometries from a small patient-specific dataset. Third, an automatic simulation data generation routine is built, enabling automatic meshing, prescribing boundary conditions, solving, and post-processing. Lastly, an efficient ML-based surrogate is developed by learning the functional map from geometries to hemodynamics in latent spaces, and the performance of using DNN and bi-fidelity technique is compared. The rest of the paper is organized as follows. The proposed framework is introduced in Section~\ref{sec:meth}. Numerical results of the surrogate modeling are presented and discussed in Section~\ref{sec:result}, and Section~\ref{sec:conclusion} concludes the paper.           

\section{Methodology}
\label{sec:meth}
The pipeline of the proposed ML-based surrogate modeling framework is summarized by Figure~\ref{fig:framework}: 
\begin{figure}[htp]
\centering
\includegraphics[width=0.6\linewidth]{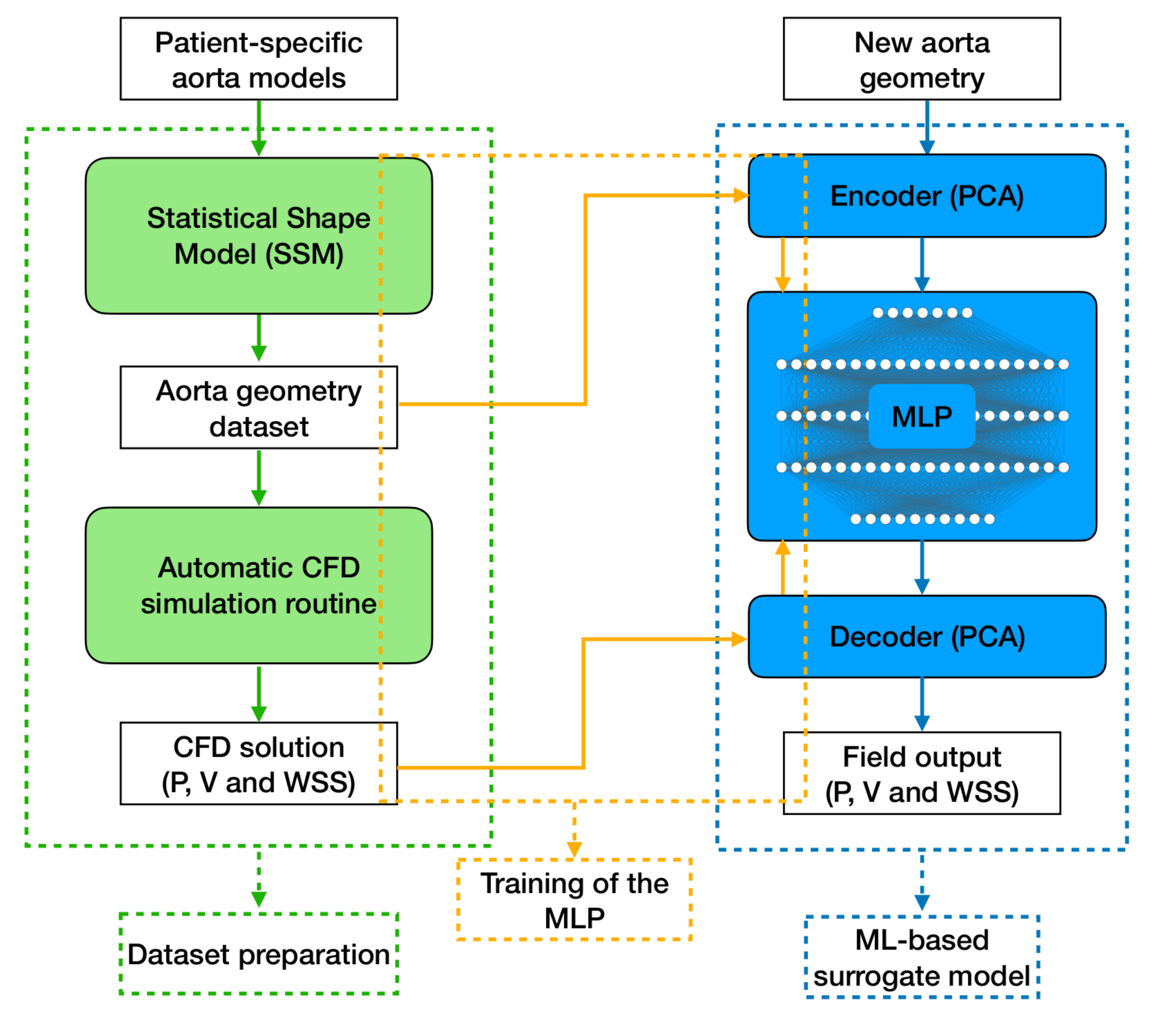} 
\caption{The schematic of the proposed ML-based surrogate modeling of patient-specific aortic flows, with several components, including dataset generation (green box), network training (orange box), and DNN inference/prediction (blue box).}
\label{fig:framework}
\end{figure}
(i) a small set of 3-D aorta geometries are reconstructed from anatomical images acquired via CTA/MRI scans on real patients; (ii) the geometric space spanned by the group of real patient-specific geometries is parameterized based on statistical shape modeling (SSM), where the variation of aorta surfaces can be described mathematically; (iii) a large number of virtual 3-D aorta shapes are synthesized within the latent geometric space; (iv) a python-based simulation routine is developed for automating meshing, equation solving, and post-processing of aortic CFD simulations on the generated ensemble of 3-D synthetic shapes, and a large dataset of hemodynamics (e.g., pressure, velocity, and WSS fields) are obtained; (v) The 3-D vascular shape and hemodynamic fields of interest (FOI) of each aorta sample are encoded into latent spaces using Principal component analysis (PCA) or proper orthogonal decomposition (POD), and then a deep neural network (DNN) is constructed to learn the non-linear mapping within the latent space. Once the network is sufficiently trained, the ML surrogate model will be able to predict the fully resolved 3-D hemodynamic flow information within a split second, given a new aorta shape.

\subsection{Patient-specific aorta geometry parameterization}
\subsubsection{Aorta dataset pre-processing}
Eight 3-D aorta geometries are reconstructed from MRI scans of patients with Coarctation of the aorta (COA) conditions by the German Heart Institute Berlin~\cite{bruning2018uncertainty}. Prior to the parameterization of the variation of aorta shapes, one crucial step is to build correspondence among them, which is very challenging since the deformation across different patients is non-isometric, especially aortas with a complex topological structure (e.g. many branches). To reduce complexity, this work is focused on 3D aortic geometries with homeomorphisum. In particular, several branches of the original thoracic aorta geometries, including left subclavian artery (LSA), left carotid artery (LCA), right subclavian artery (RSA) and right carotid artery (RCA), are trimmed off, simplifying their structure to a single channel. Later, a point cloud of 4000 vertices is sampled from each aorta using voronoi clustering implemented via the pyacvd module in python, which is loosely based on Approximated Centroidal Voronoi Diagrams (ACVD)~\cite{valetteACVD}. To regularize the position and the orientation of the aorta samples in the 3-D Cartesian coordinate system, rigid iterative closest points (ICP) algorithm~\cite{bouaziz2013sparse} is adopted to align every aorta geometry to a template selected from the aorta dataset. 

\subsubsection{Statistical shape modeling and virtual aorta synthesis}
With the aligned real aorta geometries, we establish an SSM framework to parameterize the complex 3-D patient-specific geometric input space and synthesize enormous virtual aorta geometries. The procedure is conducted in three phases: (1) building correspondence; (2) projection to latent space; (3) synthesizing virtual aortas. 

\underline{\bf Building correspondence.} Let $\{A^i\}_{i=1,...,n}$ denote the set of $n$ original patient-specific aorta shapes, and each shape sample $A^i \in \mathbb{R}^{3 \times p}$ is represented by a point cloud of surfaces vertices $A^i = \{\mathbf{a}^i_j\}_{j=1}^p$, where $p$ is the total number of vertices. Given a template geometry $A^S \in \{A_i\}_{i=1,...,n}$ selected from the set $\{A^i\}_{i=1,...,n}$, we aim to build correspondence by deforming the template $A^S$ to all the other aorta shapes $A^T \in \{A_i\}_{i\neq s}$, which are referred to as source geometry and target geometries, respectively. The diffeomprphism function $A^T = \Phi(A^S): \mathbb{R}^{3 \times p} \to \mathbb{R}^{3 \times p}$, describing the deformation from the template $A^S$ to each of the rest aorta samples $\{A_i\}_{i\neq S}$, can be solved based on the controlled-point-based large deformation diffeomorphic metric mapping (LDDMM) algorithm~\cite{beg2005computing}. In LDDMM, the deformation is described as a ``current" of surface, which is the flux of a 3-D \qq{velocity} vector field over the template $\{\mathbf{a}^S_i\}_{i = 1}^p$,
\begin{equation}
\boldsymbol{v}(\mathbf{a}_i) = \sum_{j=1}^{r} K(\mathbf{a}_i,\mathbf{q}_j) \cdot \boldsymbol{\mu}_j,
\end{equation}
where $\{\mathbf{q}_j\}_1^r$ and $\{\boldsymbol{\mu}_k\}_1^r$ represent a set of  $r$ \qq{control} points and corresponding \qq{momentum} vectors, which are defined on the source geometry $A^s$; K is a Gaussian kernel $K(\mathbf{a},\mathbf{q}) = exp(-||\mathbf{a}-\mathbf{q}||_{L_2}/\sigma^2)$ with $\sigma$ controlling the typical width of the deformation. The deformation is performed along the pseudo time axis $\tau$, and the deformed point cloud locations at $\tau$ are denoted by $A^s(\tau)$, where $A^s(t=0)$ represents the initial state (i.e., the point cloud $\{a^s_{j}\}_{j=1}^p$ of template $A_s$). The point deformations are governed by the following ODEs, 
\begin{equation}
\dot{A}^s(\tau) = \mathbf{v}\big(A^s(\tau), \tau\big),
\end{equation}
which can be solved using the second-order Runge-Kutta method, and meanwhile the control points and momentum vectors are updated based on the Hamiltonian dynamics,
\begin{subequations}
\begin{empheq}[left=\empheqlbrace]{align}
 &\dot{\mathbf{q}}(\tau) = \mathbf{K}\Big(\mathbf{q}(\tau),\mathbf{q}(\tau)\Big) \boldsymbol{\mu}(\tau), \\
 &\dot{\boldsymbol{\mu}}(\tau) =-\frac{1}{2} \nabla_\mathbf{q} {\mathbf{K}\Big(\mathbf{q}(\tau),\mathbf{q}(\tau)\Big) \boldsymbol{\mu}(\tau)^\top\boldsymbol{\mu}},
\end{empheq}
\end{subequations}
where $\mathbf{K}$ represents the kernel matrix and $\big(\mathbf{K}(\mathbf{q}(\tau),\mathbf{q}(\tau))\big)_{i,j}=K(\mathbf{q}_i(\tau),\mathbf{q}_j(\tau))$.
We force the deformed template $\tilde{A}^S_{\tau^e}$ at the final step (i.e., $\tau=\tau^e$) to match the target geometry $A^T$ using the steepest gradient descent optimization algorithm. To define the loss function, a \qq{distance} metric, evaluating the difference between the deformed template $\tilde{A}^S_{\tau^e}$ and the target geometry $A_T$, is required. As the aorta geometries are composed of triangular meshes with grid centers $\{\mathbf{c}_i\}_{i=1}^m$ and edge normal vectors $\{\mathbf{n}_i\}_{i=1}^m$, the varifold distance metric $d$ is defined as,  
\begin{equation}
d\bigg(\Big\{(\mathbf{n}^S_i,\mathbf{c}^S_i)\Big\}_{i=1}^{m^S}, \Big\{(\mathbf{n}^T_j,\mathbf{c}^T_j)\Big\}_{j=1}^{m^T}\bigg) = \sum_{i}^{m^S}\sum_{j}^{m^T}K(\mathbf{c}^S_i, \mathbf{c}^T_j)\cdot\frac{\Big((\mathbf{n}^S_i)^\top\cdot\mathbf{n}^T_j\Big)^2}{||\mathbf{n}^S_i||\cdot ||\mathbf{n}^T_j||}.
\end{equation}
The metric is calculable regardless of whether point-to-point correspondence exists between two geometries. Namely, the total amount of grids $m^S$ and $m^T$ for source and target geometries can be different. 
Once the diffeomorphism functions are solved, all target geometries can be approximated by the deformations from the template, establishing a correspondence for the original shape dataset. The LDDMM algorithm has been implemented in Deformetrica, an open-source software designed for statistical analysis of meshes~\cite{durrleman2014morphometry}.

\underline{\bf Projection to latent space.} To enable a concise parameterization of the input geometric space, PCA is applied to encode the 3-D irregular shape onto the latent space. Specifically, $\{(x^i_j,y^i_j,z^i_j)\}_{j=1,..,p}$, denoting the Cartesian coordinates of the point cloud on the aorta sample $\tilde{A}_i$, was flattened into a vector $\tilde{\pmb{V}}^i=(x^i_1,y^i_1,z^i_1,...,x^i_p,y^i_p,z^i_p)$. The entire 3-D shape dataset can be squeezed into a matrix $\pmb{A}=({\tilde{\pmb{V}}_1}^\top,...,{\tilde{\pmb{V}}_N}^\top )^\top$. Subsequently, PCA acting on $A$ yields the eigenvectors $\{\pmb{W}_i\}_{i=1,...,z}$ and eigenvalues $\{\lambda_i\}_{i=1,...,z}$, where $z$ is the truncation integer controlling the number of primary components needed. 
Each aorta shape can be decomposed as follow,
\begin{equation}
\tilde{\pmb{A}_i} \approx \pmb{A}_{mean}+\sum_{j=1}^{z}\alpha^i_j\sqrt{\lambda_j}\pmb{W}_j,
\end{equation}
where the coefficient vector $\pmb{\alpha} _i = (\alpha^i_1,...,\alpha^i_z)$ represents the encoded shape, which, in other words, is the projection of sample $A_i$ in the latent space. Let $F$ and $F^{-1}$ denote the PCA operator and its inverse, respectively. The reconstruction error $E_{pca}$ is defined to evaluate the accuracy of the PCA transform,
\begin{equation}
E_{pca} = \frac{||F^{-1}(F(\pmb{A}))-\pmb{A}||_{L2}}{||\pmb{A}||_{L2}}.
\end{equation}

\begin{figure}[htp!]
\centering
\includegraphics[width=0.8\linewidth]{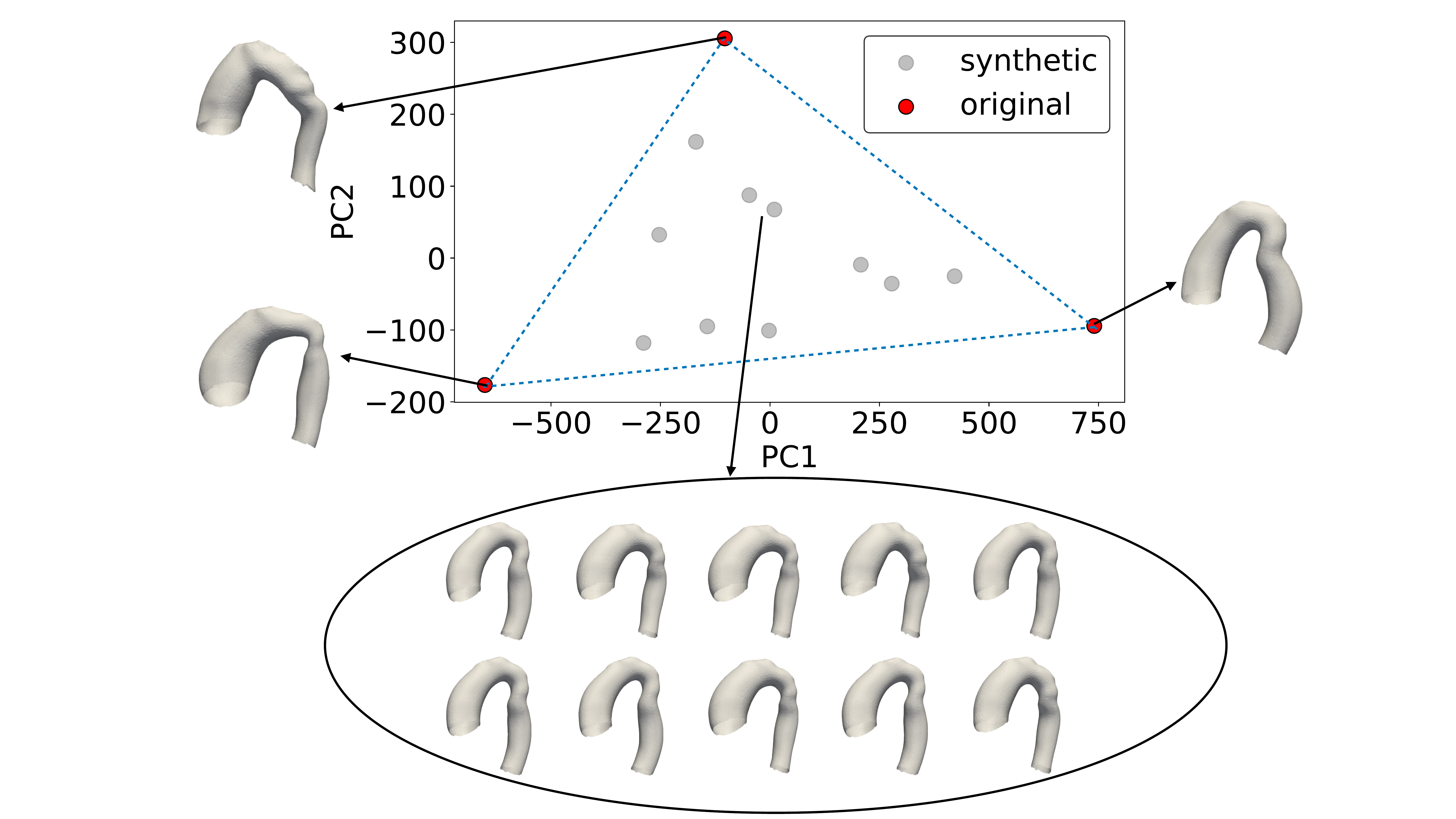} 
\caption{Illustration of latent space distribution and visualization of generated synthetic aortas (\tikzcirclegray{3pt}) from 3 original aorta shapes (\tikzcirclered{3pt}), where PC1 and PC2 represent the first and second primary components of the latent geometric space, respectively.}
\label{fig:sythesis}
\end{figure}

\underline{\bf Synthesizing virtual aorta geometries.} Motivated by the need of sufficient training samples for data-driven ML surrogate modeling, we propose a shape synthesizing method for producing a large amount of synthetic 3-D aortas from a small set of original aorta shapes. The general idea is to sample from the latent geometric space spanned by the original aorta shapes and then synthesize new aortas by decoding the sampled latent shape vectors back to the Cartesian space, as illustrated in Figure~\ref{fig:sythesis}, where the geometric space is spanned by three original aorta shapes.   
To synthesize new geometries within the geometric space, we propose two different methods: (1) random linear interpolation and (2) uniform PCA sampling. The random linear interpolation approach linearly combines the $n$ original samples in the latent space with randomly assigned weights to generate new samples $\pmb{A}^*$, 
\begin{equation}
\pmb{A}^* = F^{-1}(\sum_{i=1}^{n} \omega_i\pmb{\alpha_i}) =  \pmb{A}_{mean}+\sum_{i=1}^{n}\sum_{j=1}^{z}\omega_i \alpha^i_j\sqrt{\lambda_j}\pmb{W}_j,
\end{equation}
where $\pmb{\omega} = [\omega_1, \cdots, \omega_n]^T$ is a randomly generated weight vector that satisfies $\sum_{i=1}^{z}\omega_i = 1$. As an alternative, uniform PCA sampling approach generates new samples by uniformly perturbing the PCA coefficients within the ranges bounded by the original geometries,
\begin{equation}
\pmb{A}^* = \pmb{A}_{mean}+\sum_{j=1}^{z}\tilde{\alpha}^*_j\sqrt{\lambda_j}\pmb{W}_j,
\end{equation}
where $\pmb{\tilde{\alpha}^*} = [\tilde{\alpha}^*_1, \cdots, \tilde{\alpha}^*_z]^T$ is a randomly generated vector, and each dimension $\tilde{\alpha}^*_j$ is sampled from a uniform distribution U$\big[\min{(\alpha_j^i|_{i=1}^n)}, \max{(\alpha_j^i|_{i=1}^n)}\big]$. Figure~\ref{fig:diffeomorphism} plots a few generated samples within the 2-D latent space defined by its first (PC1) and second (PC2) primary components and the corresponding synthetic geometries. The synthetic geometries (blue) are scattered within the triangular region bounded by three original geometries (orange).

\subsection{Automatic CFD simulation routine}
To systematically generate massive CFD simulation data, we develop a python routine to automate tedious simulation procedures, including geometry pre-processing, meshing, boundary setting, simulation, and post-processing. As the first step, pre-processing of the aorta geometries is needed in order to be suitable for CFD simulations. First, since the edges at the entry and the exit of the raw aorta channel could be jaggy, the entry is slightly extended to construct a circular rim at the inlet. Similarly, the exit of the aorta is also extended, which also avoids inverse flows at the outlet. 
Second, the aorta surface is smoothed to reduce bumpiness while still preserving its geometric characteristics. Third, caps are added to seal nonphysical holes on the aorta geometry. Once the geometries pre-processing is done, 3-D unstructured triangular meshes are generated accordingly and then converted into \textit{msh} format suitable for subsequent CFD simulations. These operations are implemented 
based on the Vascular Modeling Toolkit (VMTK), which is an open-source module specialized in 3D reconstruction, geometric analysis and mesh generation. A python wrapper is implemented to automate the entire procedure in parallel for the ensemble setting.

Subsequently, a large-size ensemble of CFD simulations are performed to obtain the corresponding FOIs for each sample of the generated dataset, based on the open-source CFD platform, OpenFOAM. As handling a massive set of cases can easily be cumbersome and time-consuming, a python subroutine coupled with OpenFOAM is established, such that the procedure of each simulation, including meshing, setting up cases, specifying boundary conditions, solving, and post-processing, is automatically executed and repeated through all samples in the dataset. This process can be conducted in parallel as well. During the post-processing step, the simulated field data on CFD meshes are projected onto the surface or volumetric mesh grids where the correspondences are established. In particular, surface pressure and wall shear stress are projected onto surface vertices and velocity data are projected onto the volumetric nodes via the nearest-neighbor interpolation.   
The volumetric nodes are obtained by uniformly sampling points between the centerline and the surface vertices. Finally, a large number of input geometries and corresponding flow solutions are obtained as the labeled dataset for surrogate modeling.  

\subsection{ML-based surrogate model}
In order to build a fast forward map from the geometric space to the solution space of interest, we propose a supervised deep learning solution based on PCA encoding-decoding and fully-connected deep neural networks.

\subsubsection{Encoding-decoding formulation.} 
Considering the significantly high dimensions of 3-D shapes and solution fields in their discrete mesh forms, we adopt an encoding-decoding strategy to reduce the learning complexity. Specifically, the input geometries and output hemodynamic solution fields are encoded into low-dimensional latent spaces. To promote training efficiency, the encoder-decoder construction is decoupled with DNN-based surrogate training, and PCA (or POD) instead of DNN-based autoencoder is used for the encoding-decoding process. To balance the reconstruction accuracy and learning complexity, 100\% of the PCA total variation is preserved for the input geometries, and over 95\% is preserved for all output solution fields, resulting in reconstruction errors less than 1\%. The relationship between the encoded shape and solution fields is learned by a multilayer perceptron (MLP), where Rectified Linear Unit (Relu) is used as the activation function. 

\subsubsection{Learning architecture optimization.}
The DNN-based forward map is learned from the labeled dataset using the Adam optimizer, where the learning rate is adaptively changed based on the estimates of the moments. To achieve the best learning performance, we optimize the MLP architecture and other learning hyperparameters, including the number of layers, the number of neurons of each layer, batch size, and initial learning rate, using a Bayesian optimization algorithm. Specifically, RAY-tune~\cite{liaw2018tune}, an open-source python module for scalable hyperparameter tuning, is adopted here to achieve this goal. The general idea is to construct a posterior distribution of functions based on Gaussian processes within the prescribed ranges of the hyperparameters to be optimized. During the optimization, the posterior of hyperparameters will be updated given more and more training data, leading to the best learning configuration. The Bayesian optimization is superior to a brutal grid search method and can reach the optimized state in fewer iterations. During the optimization, the Asynchronous Successive Halving Algorithm (ASHA) is utilized to aggressively terminate non-ideal trials in advance in the interest of saving time, and more details can be found in~\cite{LiASHA}.

\subsubsection{Evaluation of performance}
To evaluate the performance of the ML-based surrogate model, relative mean square error (RMSE) is used to calculate the difference between prediction and label, which is defined as,
\begin{equation}
 RMSE = \frac{\sum_{i}^{N}(x_i - y_i)^2}{\sum_{i}^{N}{y_i}^{2}}\times100\%, 
\end{equation}
where $\pmb{X}=(x_1,...,x_N)$ and $\pmb{Y}=(y_1,...,y_N)$ are predictions and labels of FOI on $N$ vertices/grids, respectively.
For scalar fields such as pressure, RMSE is calculated directly upon the pressure field. For vector fields like wall shear stress and velocity, an array flattened from the $N\times3$ vector-matrix containing three directional components (x,y and z-direction) and the array of the vector magnitude are calculated to estimate the RMSE, referred to as vector RMSE and magnitude RMSE, respectively. When the error is evaluated upon a test dataset with multiple samples, the average of the RMSE of each sample is calculated. 

\section{Results and Discussion}
\label{sec:result}
\subsection{Statistical shape analysis and synthesis}
\begin{figure}[H]
\centering
\includegraphics[width=0.8\linewidth]{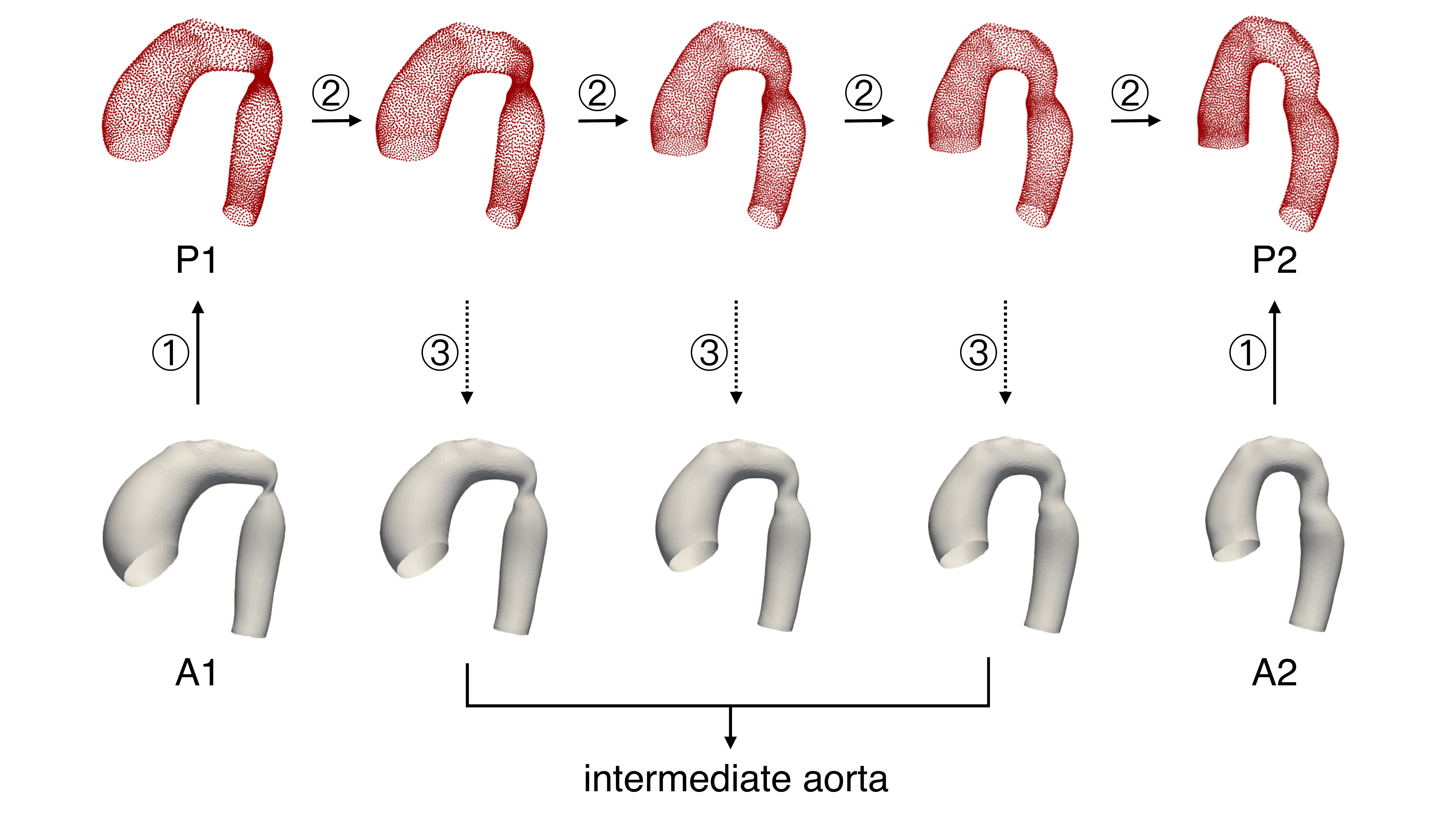} 
\caption{Building correspondence: Step\textcircled{1}, sample vertices from surface; Step\textcircled{2}, solve the diffeomorphism function; Step\textcircled{3}, reconstruct intermediate aorta}\label{fig:diffeomorphism}
\end{figure}

We first register the eight original patient-specific aorta geometries by solving seven diffeomorphisms from a randomly chosen template to the rest of seven geometries. 
An example of such diffeomorphism is demonstrated in Figure~\ref{fig:diffeomorphism}. The template $A_1$ is firstly sampled with 4000 points $P_1$ and so is the target geometry $A_2$ with $P_2$. The shape deformation $A(\tau)$ is solved with the boundary condition of $A(\tau=0) = P_1$ and $A(\tau=1) = P_2$. Several intermediate states of the function $A$ are shown in the middle, among which the point clouds near the left side resemble $P_1$ and, likewise, point clouds near the right side resemble $P_2$. In addition, the intermediate states can be used to ``interpolate" between 3-D irregular, showing geometric morphing from aorta $A_1$ to aorta $A_2$.

\begin{figure}[ht]
     \centering
     \begin{subfigure}{0.495\textwidth}
         \centering
         \includegraphics[width=\textwidth]{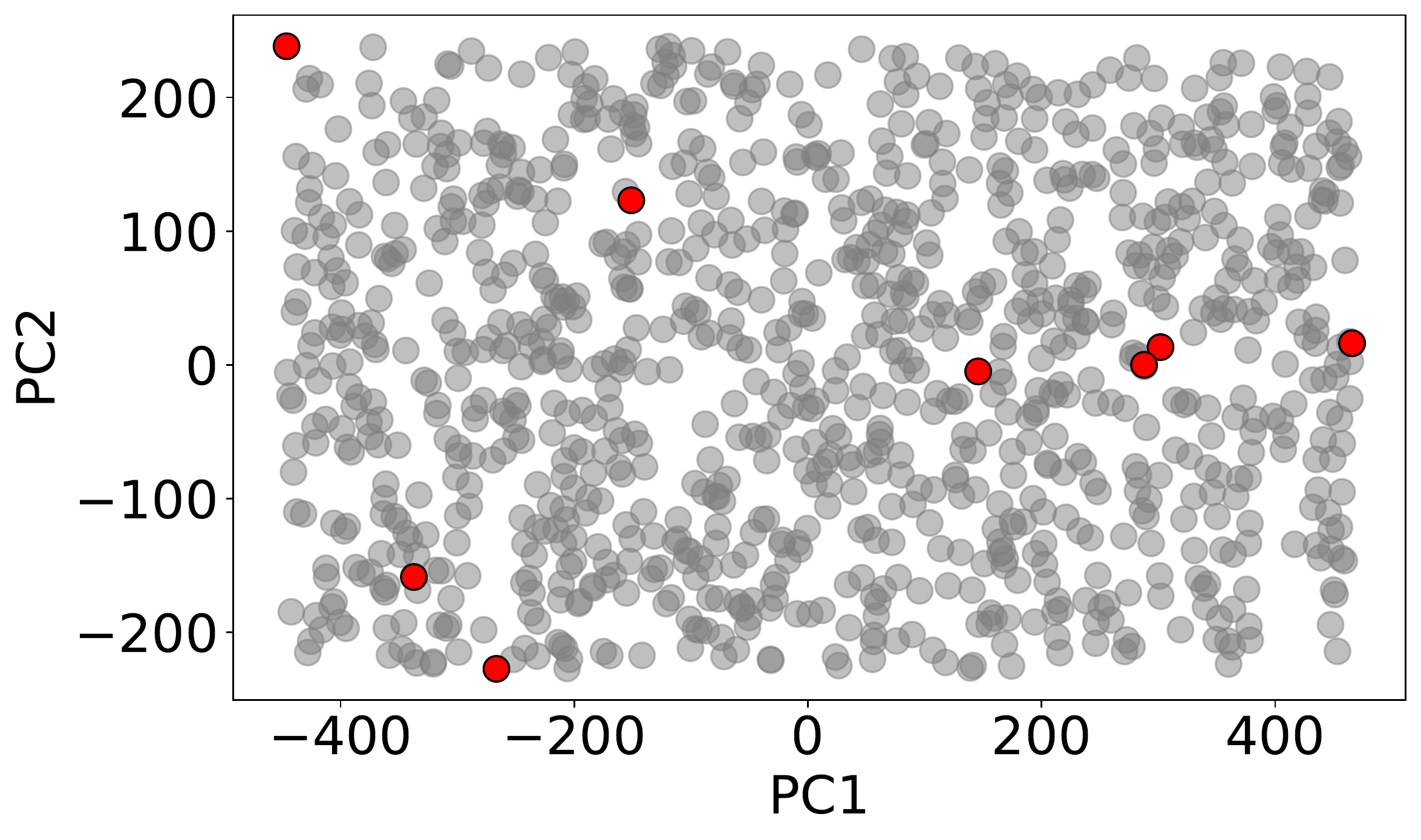}
         \caption{Uniform PCA sampling}
         \label{fig:us}
     \end{subfigure}
     \hfill
     \begin{subfigure}{0.495\textwidth}
         \centering
         \includegraphics[width=\textwidth]{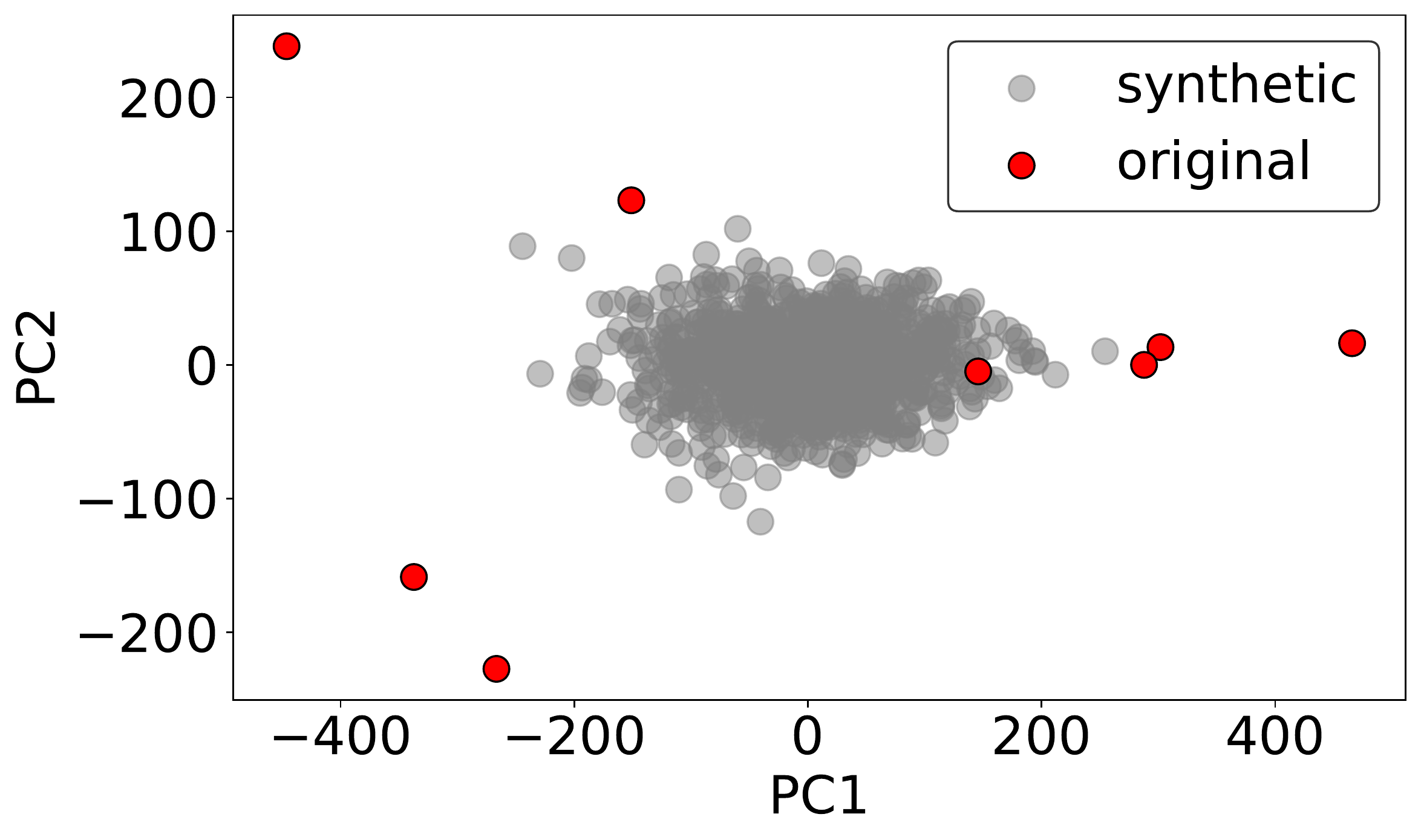}
         \caption{Random interpolation sampling}
         \label{fig:three sin x}
     \end{subfigure}
        \caption{Latent space distribution of synthetic (\tikzcirclegray{3pt}) and original (\tikzcirclered{3pt}) samples for different synthesizing method: (a) uniform PCA sampling, (b) random interpolation sampling}
        \label{fig:sample_dist}
\end{figure}
Based on the set of eight original aorta geometries, a large number of virtual aortas are synthesized by sampling the geometric space spanned by the original shapes.  
Both the uniform PCA sampling and random interpolation methods are implemented, and the sample distributions of them are plotted in Figure~\ref{fig:sample_dist} to compare their coverage of the input space. The 1,000 synthetic shapes are scattered in the coordinate plane of their PCA latent space with the first (PC1) and second (PC2) primary components.
The synthetic samples obtained by the uniform PCA sampling are much more spread out over the coordinate plane than those from the random interpolation method, indicating a better coverage of the input space and hence adopted for the data generation in this work.
\begin{figure}[ht]
     \centering
         \includegraphics[width=0.5\textwidth]{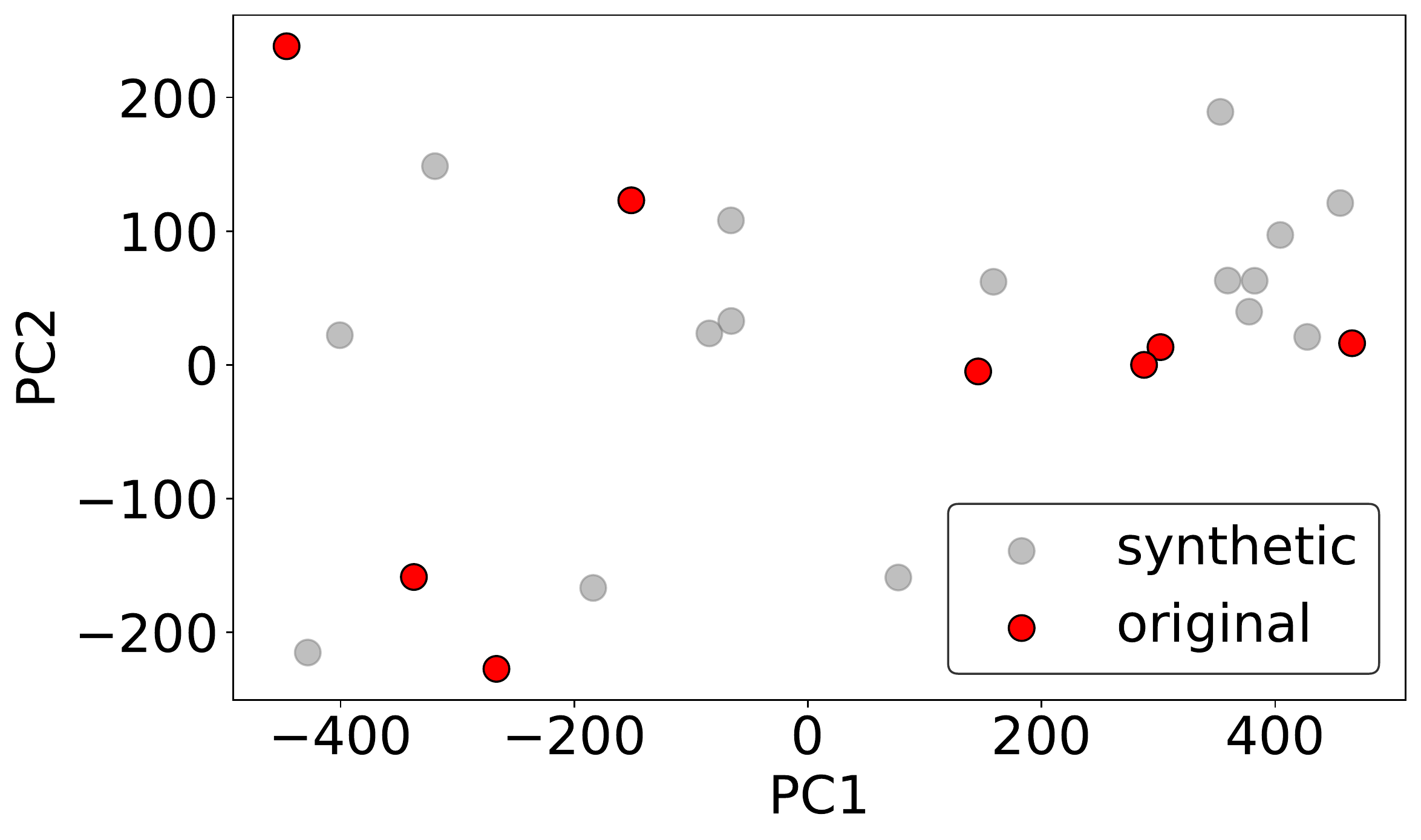}
     \hfill
         \includegraphics[width=0.9\textwidth]{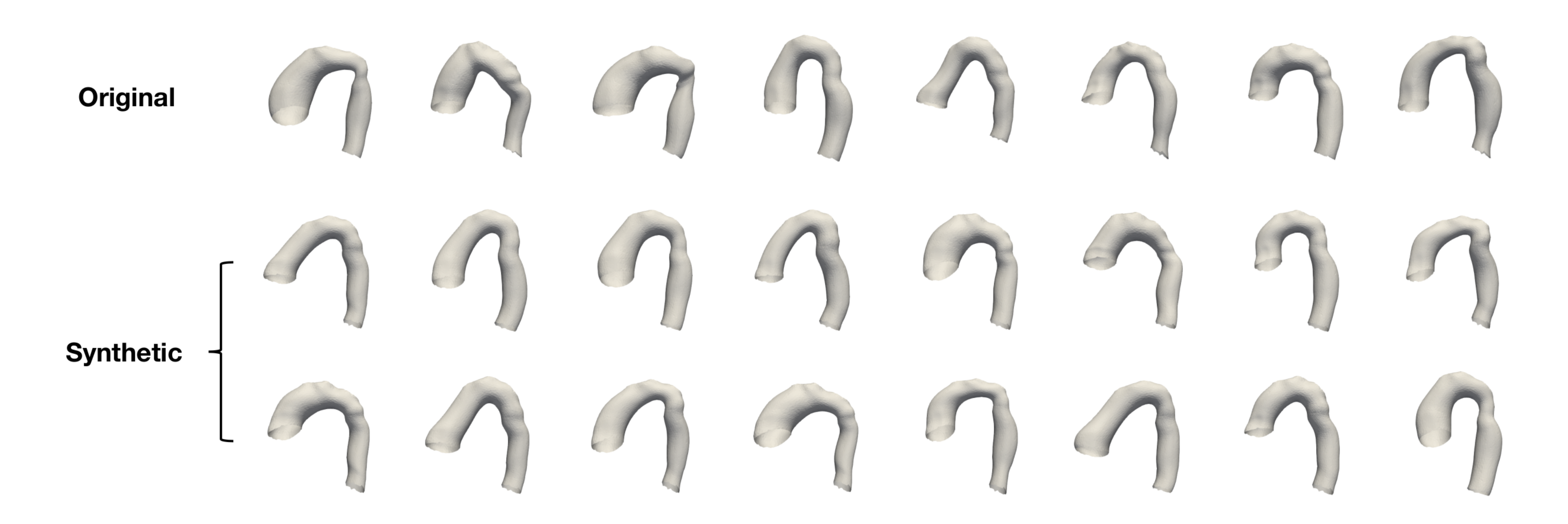}
        \caption{Latent space distribution (top) and visualization (bottom) of synthetic (\tikzcirclegray{3pt}) and original (\tikzcirclered{3pt}) aorta samples}\label{fig:sample_generate}
\end{figure}
Part of the synthetic geometries and the eight original geometries are shown in Figure 5. The uniform sampling method successfully explores the input topological space and the synthetic geometries, though virtual, possess geometric characteristics of biological shapes of real aortas and fully displays the variation of the aorta shapes. 

\subsection{Surrogate CFD modeling using deep learning}
Once the aorta geometry dataset is built, ensemble CFD simulations are conducted to generate labeled hemodynamics data, which will be used to train an ML-based surrogate. 

\underline{\bf Numerical setup.} In this work, we made the following assumptions:
1) the blood flow is Newtonian and governed by the steady-state incompressible Navier-Stokes equations;
2) the vessel walls are rigid; 
3) the inlet velocity has a parabolic profile with a maximum magnitude of 1 m/s, which is close to the peak blood flow velocity during one cardiac cycle in a human body. A zero-gradient boundary condition is applied at the inlet for pressure and a constant reference pressure is set at the outlet. 
The OpenFOAM, an open-source C++ library for FVM, is used for CFD simulations. In particular, the Semi-Implicit Method for Pressure Linked Equations (SIMPLE) algorithms~\cite{pletcher2012computational} were used for solving the incompressible Navier-Stokes equations, and the Rhie and Chow interpolation with collocated grids was employed to prevent the pressure–velocity decoupling~\cite{rhie1983numerical}. All CFD simulations were run in parallel (4 CPUs per case), and each simulation takes around $1.2\times10^3$ s to reach the convergence.
Three decoupled MLPs are constructed to predict the velocity, surface pressure, and WSS fields separately. The network architectures and learning parameters are optimized using RAY-tune module, which is given in Table~\ref{tab:nn-structure}.
The DNNs are trained on an Nvidia RTX A6000 GPU with a minimum of 2,000 epochs to ensure convergence.

\underline{\bf Learning performance study.} As a demonstration, we generated a dataset of $N_{tot} = 1,000$ aorta geometries and CFD simulated labels for training and validation study.
The dataset is divided into a training set and testing set with the ratio of $r=N_{train}/(N_{train} + N_{test})$, where $N_{train}$ and $N_{test}$ denote the size of training set and test set, respectively.
The learning-prediction performance of the trained DNN surrogates with different training-testing ratios are reported in Table~\ref{tab:nn-prediction}, where the averaged MSEs of the DNN-predicted surface pressure, velocity, and WSS fields are compared.  
Considering the randomness of the DNN initialization, the training processes for all DNNs are repeated by ten times with different random initializations, and the averaged prediction errors over these trials are reported for a more rigorous assessment.
\begin{table}[htp]
\begin{center}
\begin{tabular}{ c c c c c c} 
 \hline
 $r = \frac{N_{train}}{N_{train} + N_{test}}$ & $\epsilon_P$ & $\epsilon_V$ & $\epsilon_{VM}$ & $\epsilon_{WSS}$ & $\epsilon_{WSSM}$\\
 \hline
 r=60\% & 1.83\% & 0.735\% & 0.504\% & 1.88\% & 1.33\% \\
 r=70\% & 1.81\% & 0.726\% & 0.5\% & 1.86\% & 1.31\% \\
 r=80\% & 1.67\% & 0.734\% & 0.505\% & 1.86\% & 1.32\% \\
 r=90\% & 1.48\% & 0.64\% & 0.434\% & 1.74\% & 1.26\% \\
 \hline
\end{tabular}
\caption{DNN prediction errors (averaged MSE) with different training to testing size ratios for hemodynamics fields of interest, including surface pressure ($\epsilon_P$), velocity vector ($\epsilon_V$), velocity magnitude ($\epsilon_{VM}$), wall shear stress ($\epsilon_W$), wall shear stress magnitude ($\epsilon_{WM}$). }\label{tab:nn-prediction}
\end{center}
\end{table}\vspace{-1.5em}
It can be seen that the relative errors of the NN surrogate prediction are very low once the DNNs are sufficiently trained. For velocity field prediction, the prediction error is less than $1\%$ even only $60\%$ of the dataset is used for training. Compared to the velocity, the errors of surface field predictions such as surface pressure and WSS fields are slightly higher, since they are more sensitive to the shape of vessel walls and the boundary layer conditions, which are not easy to capture well. 
Also, for vector-valued fluid quantities like velocity and WSS, the magnitude errors are smaller than the vector errors by about $30\%$. 
Viewing the columns for each flow quantity, the prediction accuracy increases as the training size grows, which is as expected.  
For example, the pressure error climbs from $1.48 \%$ at $r=90\%$ to $1.83 \%$ at $r=60 \%$ as $r$ grows. In general, the prediction of the FOIs is accurate, indicating the great capability of the proposed DNN-based surrogate model.

\begin{figure}[H]
\centering
\includegraphics[width=0.8\linewidth]{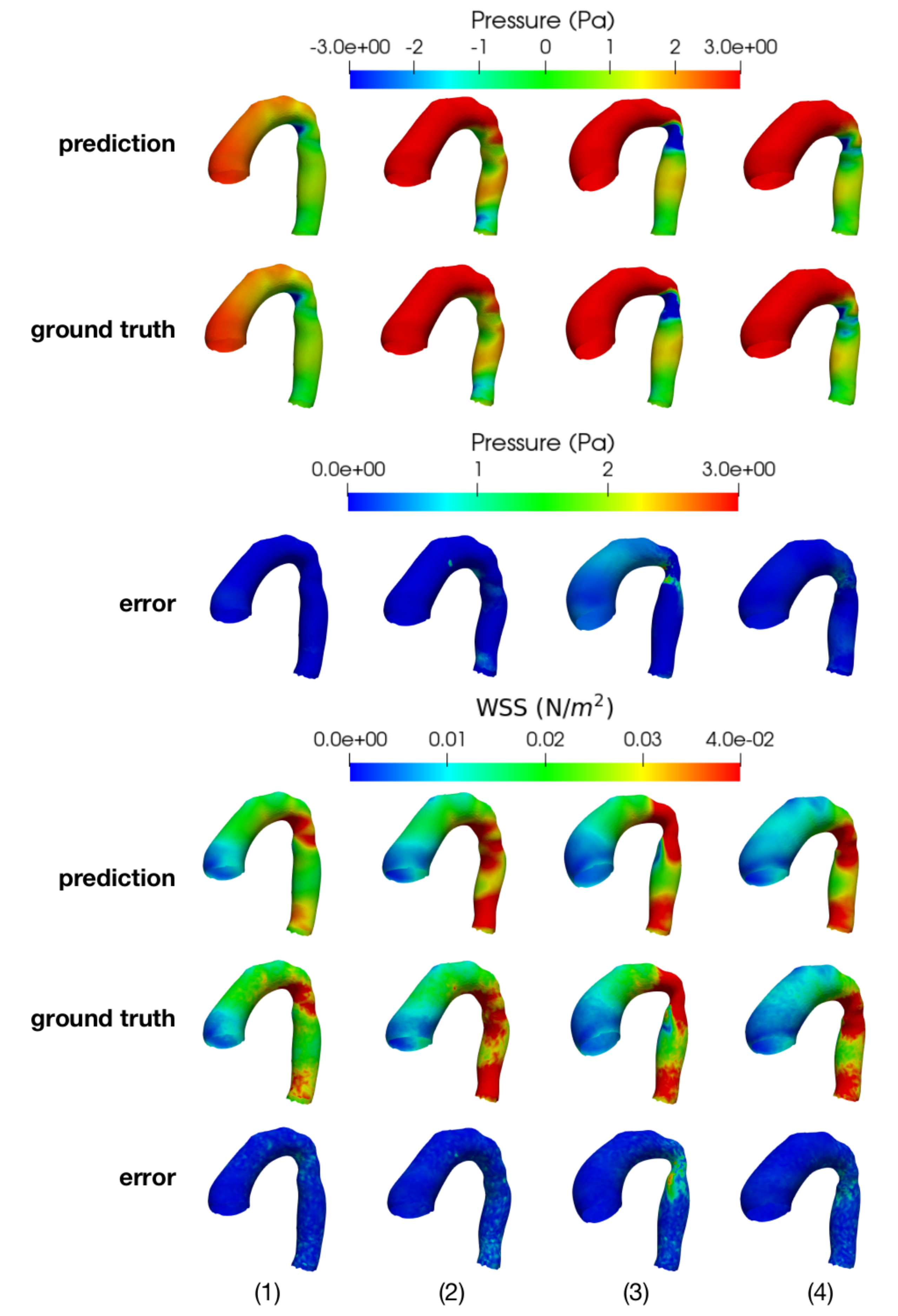} 
\caption{ML surrogate predictions of pressure (top) and wall shear stress (bottom) are compared with the CFD ground truth: prediction (1st and 4th row), CFD ground truth (2nd and 5th row), prediction error (3rd and 6th row)}\label{fig:pressure}
\end{figure}

\begin{figure}[H]
\centering
\includegraphics[width=0.8\linewidth]{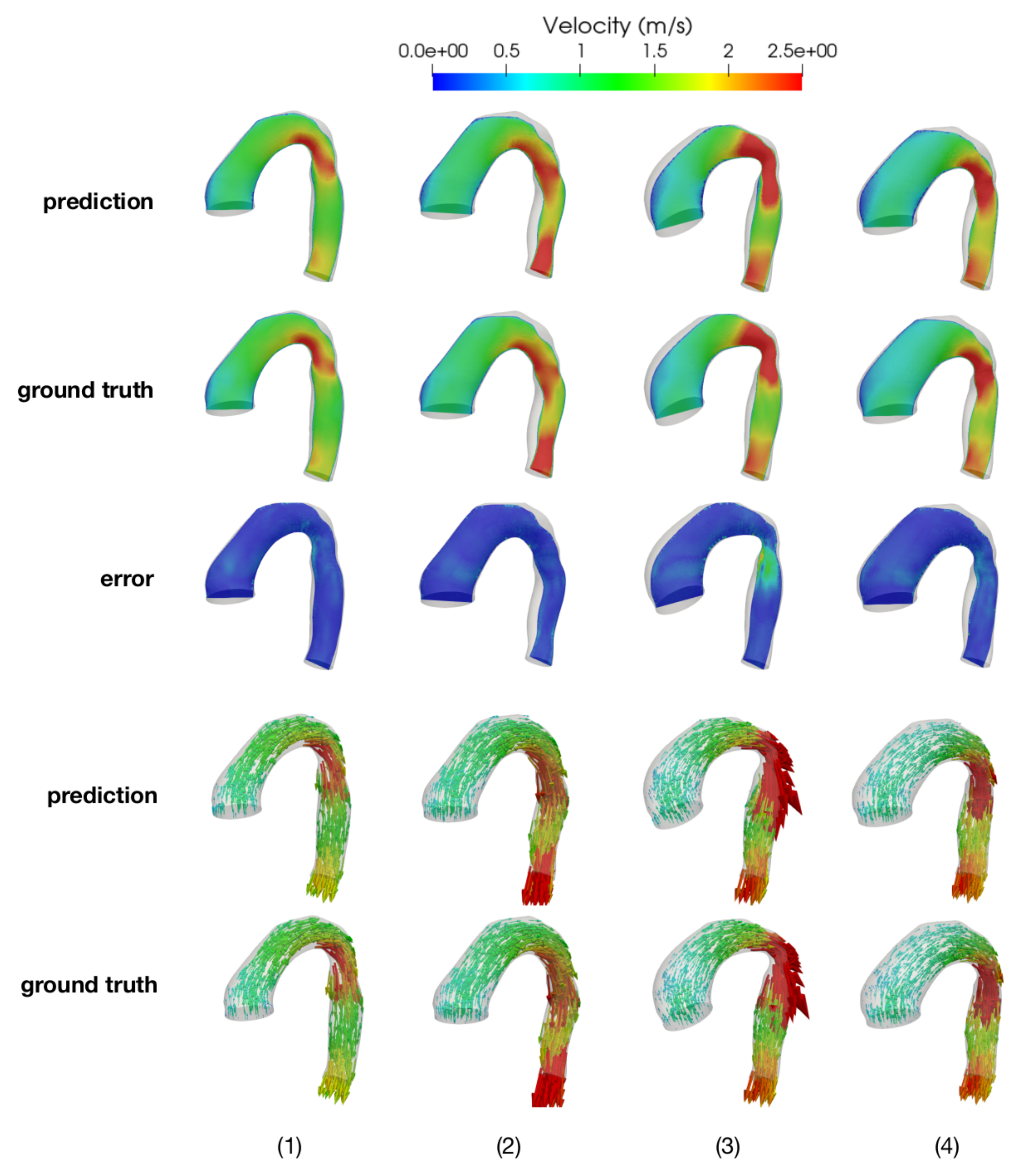} 
\caption{Cross-section velocity magnitude (top) and velocity vector filed (bottom) comparison of prediction and CFD ground truth: prediction (1st and 4th row), ground truth (2nd and 5th row), error (3rd row)}\label{fig:velocity}
\end{figure}

\underline{\bf Predicting local hemodynamic information.} In contrast to traditional surrogate model (e.g., GP models) or reduced-space model (e.g., LP/1D models), the proposed ML-based surrogate model is able to rapidly predict local hemodynamic information such as blood flow pattern, surface pressure and WSS distributions over the vessel walls. For illustration, we randomly selected four test aorta geometries, and the hemodynamic fields predicted by the DNN surrogate ($r = 80\%$) are compared with the CFD reference (referred to as ground truth), as shown in Figures~\ref{fig:pressure} and~\ref{fig:velocity}. 
Besides, the absolute error contours are also plotted out by substracting the ground truth from the prediction. 
For pressures, the surrogate model yields very close surface distribution to the ground truth, yet slight differences exist at the stenosis of the descending aorta. From the error contours, this discrepancy is very mild on samples 1, 2 and 5, but notable on sample 3, where the severity of stenosis is much higher than other samples. This indicates that the surrogate model is slightly less accurate in capturing flow features with large pressure gradients, since more flow vortices are induced by sharp changes of surface curvatures complicating the flow physics.  
Surrogate prediction of the WSS distribution is in a good agreement with the ground truth. The pattern and low/high WSS regions can be accurately captured. However, the surrogate model tends to smooth out the small WSS fluctuations compared to the CFD reference. It appears that high-frequency components of the output are partially trimmed off when interpolating among neighboring training samples based on the non-linear relations learned by the neural network. Similar to pressure, notable inconsistency is also observed near the stenosis on the descending aorta for sample 3. The velocity contours on the cross-section and internal velocity vector fields are visualized in Figure~\ref{fig:velocity}. 
Again, the predicted velocity cross-section contours agree with the CFD ground truth very well and the overall velocity vector fields from the neural network and CFD are almost identical for all samples, though the error contours of planar velocity show slight prediction errors near the stenosis region for Sample 3. 

\underline{\bf Training cost and prediction speedup.} The mappings between shape input and FOIs, including pressure, velocity and WSS, are learned using separated neural networks, and each takes about two minutes to be fully trained on an Nvidia RTX-3090 GPU. The network inference for solution field prediction takes less than 0.0012 seconds. In contrast, a full-resolution CFD simulation needs about 20 minutes to reach the steady state, not counting the overhead for mesh generation and case setup, which could also be time-consuming. As a result, the fully trained DNN surrogate model has $100,000$x speedup over the full CFD simulation in the online inference phase. We should note that a fair comparison should also consider the offline cost of constructing the surrogate, which includes the cost for network training and label generation. As for network training, the cost is negligible compared to that of a CFD simulation, thanks to the light DNN structure defined in the latent space. However, the cost for data generation depends on the number of training labels required, which is determined by the desired trade-off between the efficiency and accuracy of the surrogate. For example, supposing 500 CFD labels are used for network training, the CFD data generation cost is $500\times20$ minutes, which is significantly greater than that of a single CFD simulation. Nonetheless, the surrogate model is built for many-query applications (e.g., optimization, uncertainty quantification, inverse modeling, etc.), and the huge speedup of single model evaluation can be leveraged when a large number of model evaluations are required as the offline cost is paid off at once. For example, Markov chain Monte Carlo (MCMC), as the gold standard Bayesian inference method for forward and inverse UQ, usually requires hundreds of thousands of model evaluations to reach convergence, which well justifies the data generation at the cost of hundreds of full-order CFD evaluations. Moreover, it is worth noting that the training dataset size is application-dependent and the data generation cost can be significantly reduced if one allocates the computational budget wisely. A comprehensive study on the use of the proposed surrogate model in specific UQ or optimization applications is out of the scope of this work and will be conducted in the future.   

\subsection{Comparison of DNN-based and bi-fidelity surrogate models}
Lastly, we investigate how well the proposed DNN-based surrogate model performs compared to other state-of-art surrogate models that can provide full-field hemodynamic predictions. To enable the full-field prediction capability with the same resolution as that of full-order CFD simulations, a multi-fidelity strategy is often adopted for surrogate modeling. The multi-fidelity paradigm is pioneered by Kennedy \& O'Hagan~\cite{KennedyMF} decades ago, originally from a statistical point of view (e.g., GP-based multi-model approach). Very recently, multi-fidelity strategy has been applied for surrogate modeling in inference and uncertainty quantification (UQ) problems in cardiovascular biomechanics problems~\cite{biehler2015towards,biehler2017probabilistic,fleeter2020multilevel,gao2021bi,gao2020bi1}. In particular, Gao et al.~\cite{gao2021bi} developed a bi-fidelity (BF) surrogate modeling approach for 3-D patient-specific hemodynamic simulations, which has been demonstrated effective for both forward and inverse UQ problems~\cite{gao2021bi,gao2020bi1}. The theoretical backbone of the BF surrogate is based on the multi-fidelity paradigm,
which leverages the accuracy of high-fidelity CFD solutions and the efficiency of low-fidelity CFD solutions, largely reducing the total computation cost compared with conventional full-order CFD simulations. 
Here we compared the proposed DNN-based surrogate with the state-of-the-art BF surrogate model, which are built with the same amount of training labels. It is worth noting that the training overhead and prediction cost of the BF surrogate are always higher than those of the DNN-based surrogate with the same amount of training data (i.e., high-fidelity CFD labels), because low-fidelity CFD simulations have to be performed for both training and inference. Specifically, to build the BF surrogate model, in addition to high-fidelity (HF) CFD training data, another 1,000 low-fidelity (LF) CFD simulations are conducted on low-resolution meshes. As the low-res meshes are very coarse and the convergence criterion is set to be large, the LF simulation cost is much lower than that of a HF simulation. The LF data are used to explore the input geometric space to determine important points, where HF CFD simulations are performed for training. More details of the BF surrogate construction can be found in~\cite{gao2021bi}.      

\begin{figure}[ht]
     \centering
     \begin{subfigure}{0.32\textwidth}
         \centering
         \includegraphics[width=\textwidth]{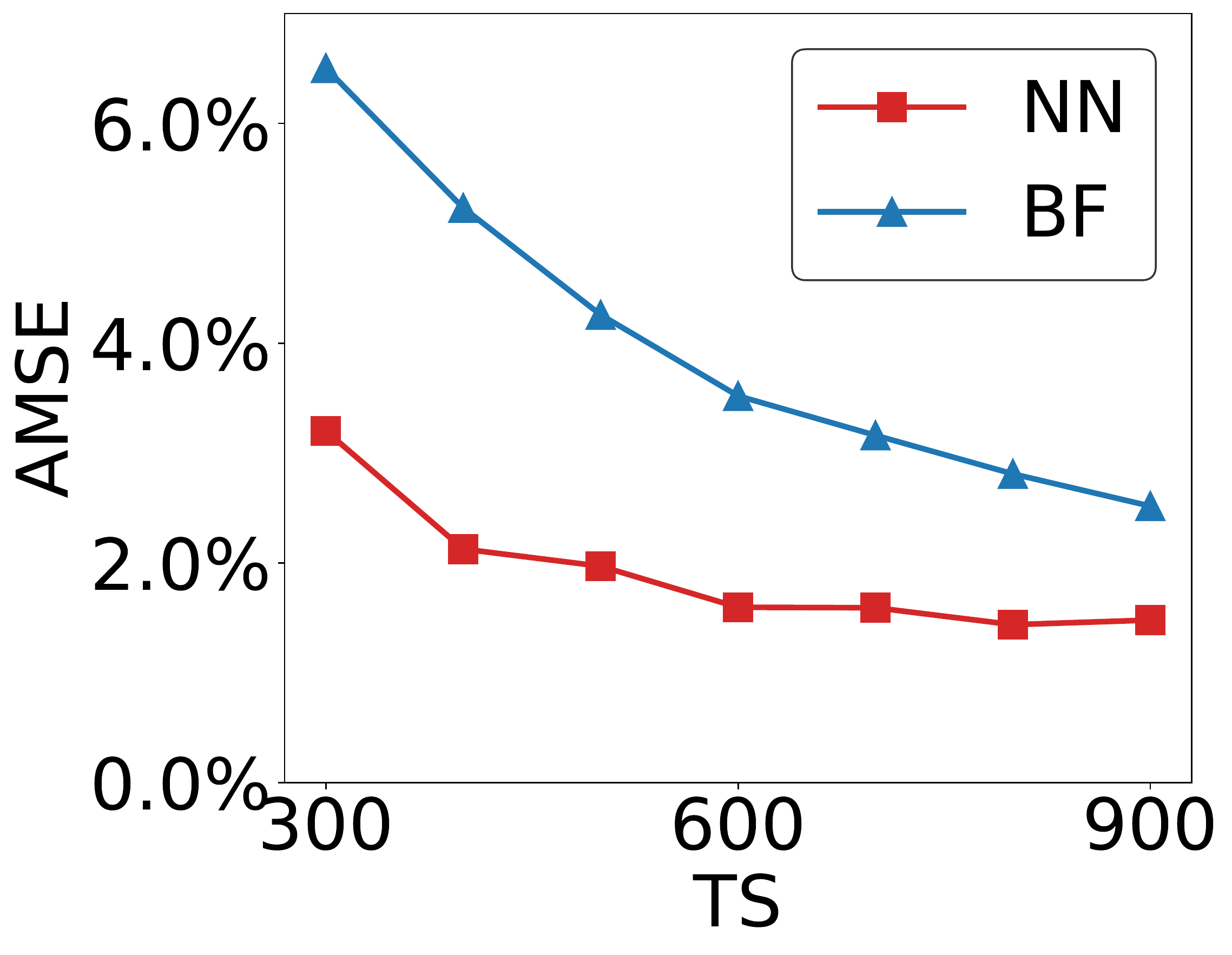}
         \caption{Pressure}
         \label{fig:psen}
     \end{subfigure}
     \hfill
     \begin{subfigure}{0.32\textwidth}
         \centering
         \includegraphics[width=\textwidth]{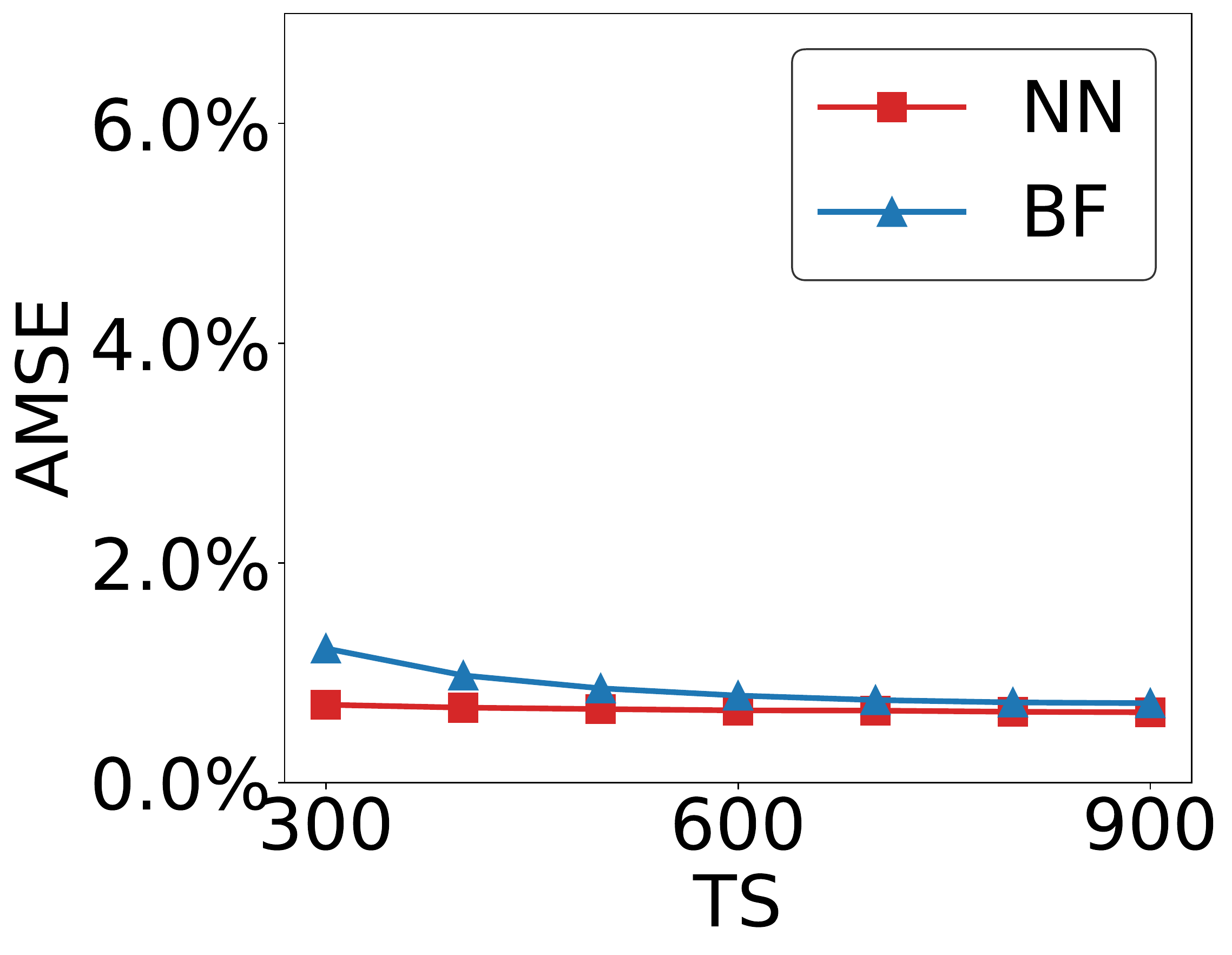}
         \caption{Velocity}
         \label{fig:usen}
     \end{subfigure}
\hfill
     \begin{subfigure}{0.32\textwidth}
         \centering
         \includegraphics[width=\textwidth]{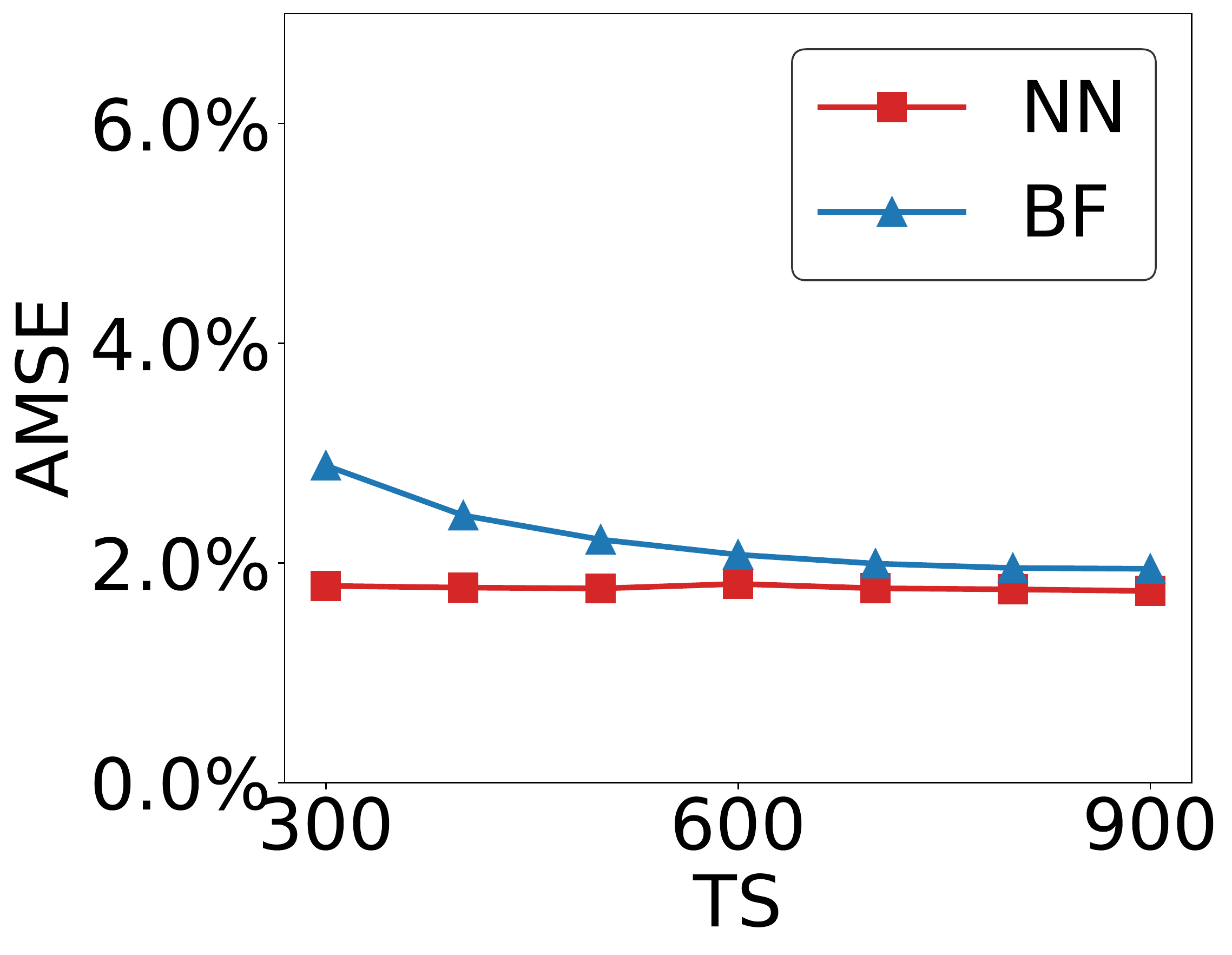}
         \caption{Wall shear stress}
         \label{fig:wsen}
     \end{subfigure}
        \caption{Comparison of neural network(NN) and bi-fidelity model(BF): Y axis, averaged mean square error (AMSE); X axis, trainset size (TS)}
        \label{fig:puwsen}
\end{figure}
To find out how the prediction errors change with the training size, the training set size (TS) changes from 300 to 900, and the testing set is a fixed group of 100 aortas randomly generated. The comparison of performance is shown in Figure~\ref{fig:puwsen}. Both surrogate models have reasonably good predictions, but the proposed DNN-based surrogate model outperforms the BF surrogate model for all FOIs under all different training scenarios. Both surrogate models reach their best accuracy at $TS=900$ for all FOIs. The prediction error of the BF model quickly increases as the training sample size decreases, while this trend is less notable for the DNN-based model, particularly for velocity and WSS predictions, indicating a more robust behavior in the small data regime. For both models, the learning performance for the 3-D velocity vector field is slightly better than that of surface field outputs, such as pressure and WSS, due to the fact that they are more sensitive to the local changes of the input geometries cross difference patients. Moreover, the BF prediction cost is higher than the DNN-based surrogate, since each BF prediction requires an LF CFD simulation. Therefore, the proposed DNN-based surrogate model is superior to the BF model in terms of accuracy and efficiency, though both models can predict all FOIs reasonably well.


\section{Conclusion}
\label{sec:conclusion}
In this study, we developed a statistical shape modeling (SSM) approach to parameterize the geometric space spanned by a set of 3-D patient aorta geometries and systematically synthesize a large number of virtual aorta geometries.  
The results show that the proposed SSM method successfully established correspondence among aorta geometries from different patients and the uniform PCA sampling algorithm sufficiently explored the input space, generating abundant synthetic aorta shapes that fully cover the geometric variation. Subsequently, an ML-based surrogate model is proposed to predict comprehensive local hemodynamic information (e.g., surface pressure, velocity and WSS fields), where an encoding-decoding structure is built based on PCA and multiple MLPs are constructed to learn the non-linear mappings between the input geometries and flow fields of interest within the latent spaces. The PCA encoder and decoders managed to reduce the dimension of the mesh-based 3-D aorta shapes and field outputs to a large extent and maintained a reconstruction error as low as $1\%$. The MLPs are well optimized in terms of their hyperparameters and trained on different sizes of sample sets. The testing results prove that the proposed surrogate model is able to yield hemodynamic field predictions in a good agreement with the ground truth obtained from full-order CFD simulations, yet the accuracy is slightly lower near the stenosis of the descending aorta, where flow features are more complicated due to sharp pressure gradients.
The accuracy of the proposed model can be improved by increasing the size of training set. The proposed DNN-based surrogate is compared with the state-of-the-art bi-fidelity (BF) surrogate model, and the DNN surrogate outperforms the BF surrogate in terms of both efficiency and accuracy for all FOIs under different training scenarios. 
Furthermore, the proposed model exhibits a robust behavior when the number of samples drops whereas the BF model easily loses its accuracy, especially for surface pressure predictions. In conclusion, the proposed DNN-based surrogate model shows a great capability of approximating the non-linearity of the CFD simulations for cardiovascular flows. Considering extremely low costs of DNN for inference, the trained DNN-based surrogate is able to provide high-resolution local hemodynamic information in split of second, showing a great potential for applications requires massive model queries, such model inference, optimization, and uncertainty quantification problems. Although the current work focus on aortic flows, the proposed methods can be applied to hemodynamic modeling for other anatomies in general. 


\section*{Acknowledgment}
The authors would like to acknowledge the funds from National Science Foundation under award numbers CMMI-1934300 and OAC-2047127, and startup funds from the College of Engineering at University of Notre Dame in supporting this study. 

\section*{Compliance with Ethical Standards}
Conflict of Interest: The authors declare that they have no conflict of interest.

\appendix
\section{Optimized MLP structures and hyperparameters}
\renewcommand{\arraystretch}{0.7}
\begin{table}[H]
\begin{center}
\begin{tabular}{c | c c c c }
 \hline
 \pmb{FOI} & \pmb{DNN Layers}       & \pmb{In \& Out features} & \pmb{batch size} & \pmb{learning rate} \\
 \hline
        & Linear Layer & (7,64)  &   &   \\
        & Relu          & -    &   &                                        \\
        & Linear Layer & (64,8)  &   &       \\
Pressure & Relu          & -   & 8  & $1\times10^{-3}$                                         \\
        & Linear Layer & (8,256)  &   &     \\
        & Relu          & -   &   &                                        \\
        & Linear Layer & (265,10)   &   &    \\
 \hline
         & Linear Layer & (7,64)  &   &   \\
        & Relu          & -    &   &                                        \\
        & Linear Layer & (64,512)  &   &       \\
Velocity & Relu          & -   & 8  & $4.07\times10^{-4}$                                         \\
        & Linear Layer & (512,256)  &   &     \\
        & Relu          & -   &   &                                        \\
        & Linear Layer & (265,10)   &   &    \\
 \hline
         & Linear Layer & (7,512)  &   &   \\
        & Relu          & -    &   &                                        \\
        & Linear Layer & (512,16)  &   &       \\
      & Relu          & -   &    &                                          \\
        & Linear Layer & (16,512)  &   &     \\
        & Relu          & -   &   &                                        \\
WSS     & Linear Layer & (512,32)   & 8  & $2.30\times10^{-4}$   \\
        & Relu          & -   &   &                                        \\
        & Linear Layer & (32,16)   &   &    \\
        & Relu          & -   &   &                                        \\
        & Linear Layer & (16,256)   &   &    \\
        & Relu          & -   &   &                                        \\
        & Linear Layer & (256,10)   &   &    \\
 \hline
\end{tabular}
\caption{The optimized MLP structure and hyperparameters based on Bayesian tuning}\label{tab:nn-structure}
\end{center}
\end{table}




\bibliographystyle{elsarticle-num}
\bibliography{ref}

\end{document}